# A Novel Energy-Efficient Salicide-Enhanced Tunnel Device Technology Based on 300 mm Foundry Platform Towards AIoT Applications

Kaifeng Wang[1], Qianqian Huang[1,4,5*], Yongqin Wu[2], Ye Ren[2], Renjie Wei[1], Zhixuan Wang[1], Libo Yang[1], Fangxing Zhang[3], Kexing Geng[3], Yiqing Li[1], Mengxuan Yang[1], Jin Luo[1], Ying Liu[1], Kai Zheng[2], Jin Kang[2], Le Ye[1], Lining Zhang[3], Weihai Bu[2*] and Ru Huang[1,4,5*]
[1]School of Integrated Circuits, Peking University, Beijing 100871, China (*E-mail: hqq@pku.edu.cn, ruhuang@pku.edu.cn); [2]Semiconductor Technology Innovation Center (Beijing), Beijing 100176, China (*E-mail: WeiHai_Bu@sticbj.com); [3]School of Electronic and Computer Engineering, Peking University, Shenzhen 518055, China; [4]Chinese Institute for Brain Research, Beijing, Beijing 102206, China; [5]Beijing Advanced Innovation Center for Integrated Circuits, Beijing 100871, China.

***Abstract*—This work demonstrates a novel energy-efficient tunnel FET (TFET)-CMOS hybrid foundry platform for ultra-low-power AIoT applications. By utilizing the proposed monolithic integration process, the novel complementary n- and p-type Si TFET technology with dopant segregated source junction and self-aligned drain underlap design is successfully integrated into a 300mm CMOS baseline process without CMOS performance penalty and any new materials, experimentally demonstrating the large $I_{ON}$ and record high $I_{ON}/I_{OFF}$ ratio of $10^7$ among TFETs by industry-manufacturers. The device performance and variability are also co-optimized for high-volume production. Further circuit-level implementation are presented based on the calibrated compact model. The proposed TFET-CMOS hybrid logic and SRAM topologies show significant energy efficiency improvement with comparable operation speed compared with standard CMOS circuit, indicating its great potential for power-constraint AIoT applications.**

## I. INTRODUCTION

The fast expanding market of Artificial Intelligence of Things (AIoT), which is being considered as one of the most promising, disruptive and cutting-edge driving force for AI applications, has stimulated a lot of innovations from device to circuit level [1-2]. Despite of diverse requirements from different AIoT chips, low power consumption is always the first priority [3-4]. Tunnel FET (TFET) has already demonstrated its superiority in terms of power consumption among emerging device technologies, while suffering from the fundamental low on-state current ($I_{ON}$) issue [5]. Nevertheless, it can be expected that by using TFET-CMOS hybrid system, such as co-integrating a TFET-based intelligent wake-up chip with a CMOS-based high performance system (Fig. 1), both ultra-low power, high accuracy and real-time can be achieved for AIoT nodes in random-sparse-event (RSE) scenarios. So far, many industry-level manufacturers has demonstrated lots of novel TFET structures to improve the device $I_{ON}$ [6], while the device $I_{ON}/I_{OFF}$ ratio and CMOS compatibility are still poor for high-volume production. For Si TFETs, the $I_{ON}$ is usually less than 1μA/μm, which cannot satisfy the real time requirement of AIoT applications. For TFETs with new materials like III-V, although the device $I_{ON}$ can be effectively enhanced, they suffer from the high-static-energy-consumption issue which is more concerned for AIoT. Moreover, the TFET device technology also face challenges like poor device complementarity [7] and large device variation [8]. For circuit, the features of TFETs, such as uni-directional

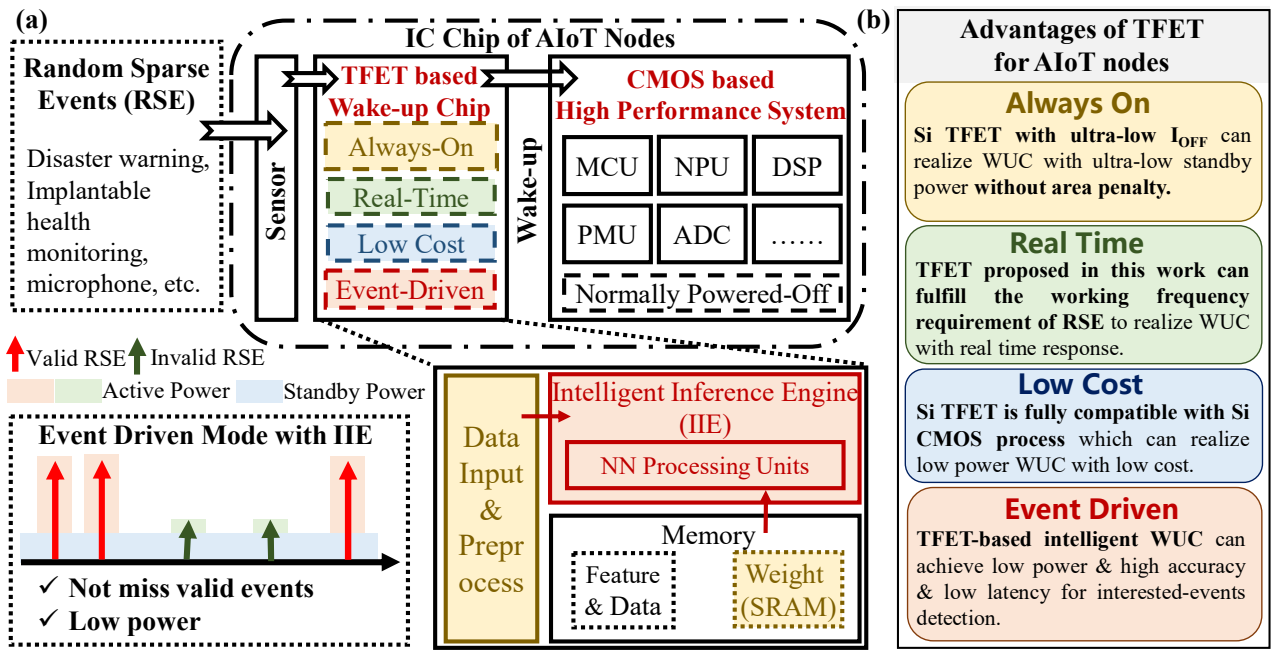

Fig.1 (a) A TFET-based intelligent wake-up chip (WUC) co-integrated with a CMOS-based high performance system (HPS). (b) The advantages of TFET for AIoT nodes with always on, real time, low cost and event driven requirements.

conduction of tunnel current [9], forward p-i-n current [10], ambipolar current [11], and TFET stacking problem [12], may make conventional circuits topologies no longer applicable.

In this work, we proposed a novel complementary dopant-segregated TFET (DS-TFET) structure co-integrated with baseline CMOS technology in 12-inch foundry platform, and experimentally demonstrated the record $I_{ON}/I_{OFF}$ ratio without CMOS device performance penalty. The monolithic integration of TFET and CMOS makes it possible to take advantage of low-power TFET and high-performance CMOS simultaneously. The proposed hybrid TFET-CMOS logic cells and SRAM are further presented with both much lower power consumption and comparable operation speed than CMOS circuits, which is very promising for realizing comprehensive requirements of AIoT nodes applications.

## II. C-DS-TFET STRUCTURE AND DESIGN OPTIMIZATION

In our previous work, we has successfully manufactured conventional Si TFET device structure using 0.13μm CMOS baseline platform [8]. Although it has demonstrate superior compatibility with standard MOSFET technology, due to the ion implantation limitation in improving junction abruptness and the relatively large gate oxide thickness, the device $I_{ON}/I_{OFF}$ ratio can still not satisfy the AIoT requirements. In this work, to further optimize TFET performance, the fabrication is based on the standard 55nm low-leakage CMOS foundry platform which is more commonly utilized towards low-cost and low-power AIoT applications, and a novel device structure called DS-TFET with asymmetric sidewalls and self-aligned dopant segregation junction is

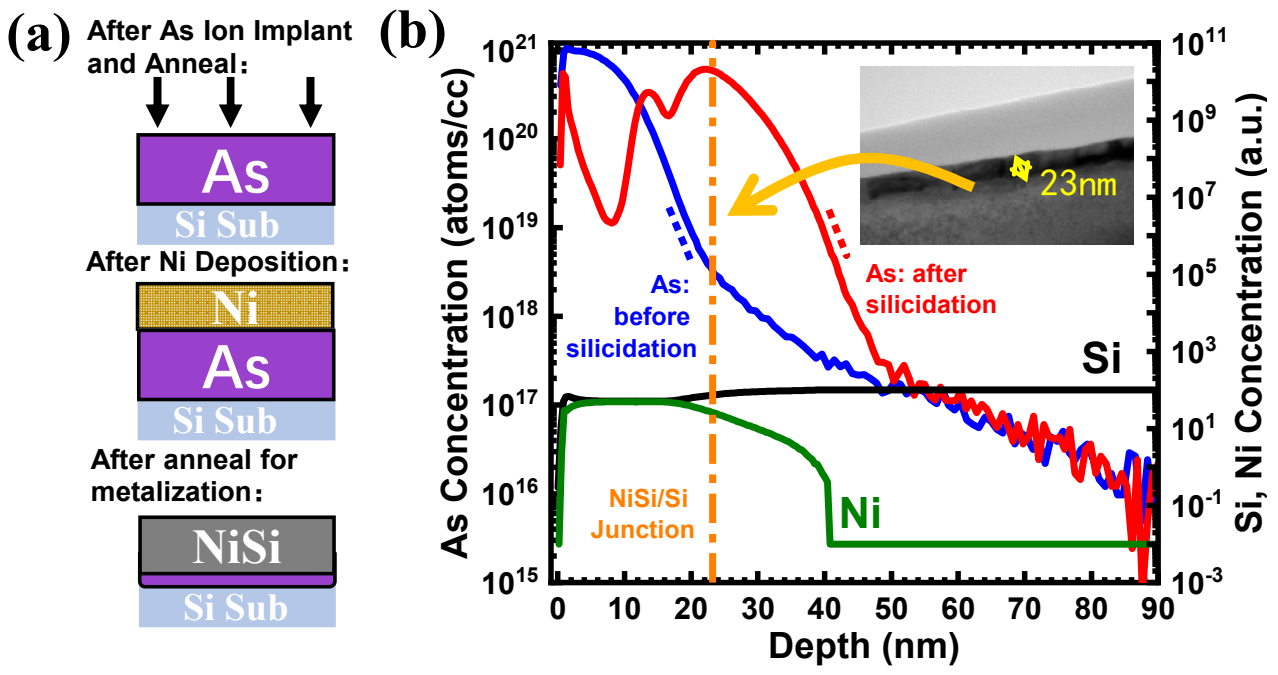

Fig.2 (a) Schematic structures and fabrication process flow of dopant segregation junction; (b) TEM and SIMS results indicate dopant segregation with more abrupt doping profile after silicide process.

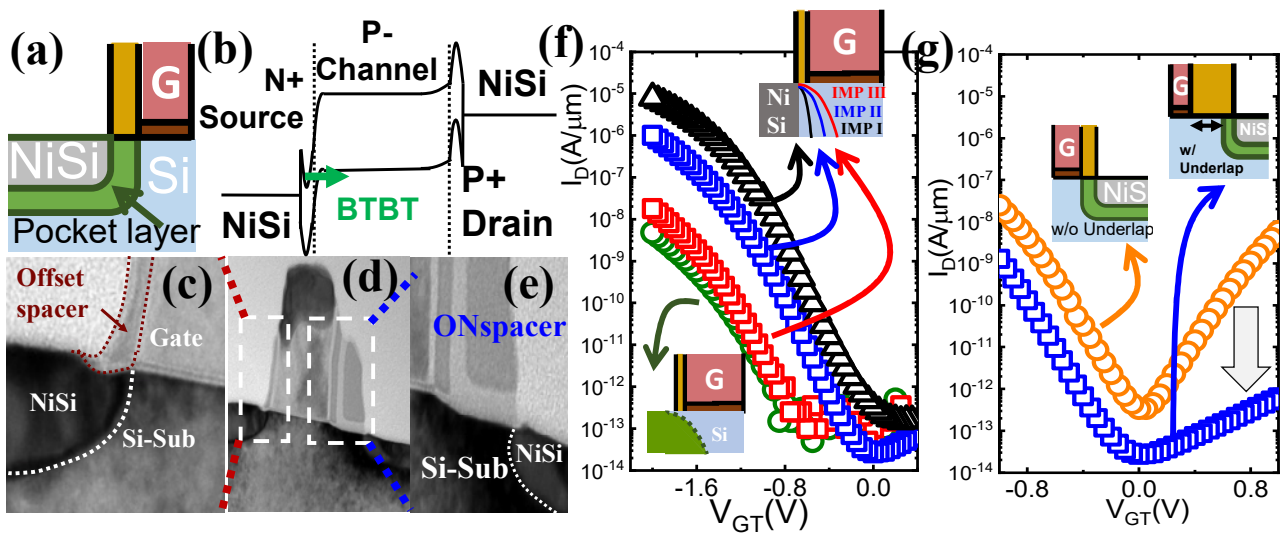

Fig.3 (a) Schematic self-aligned segregation source junction; (b) Schematic surface energy band of pDS-TFET when BTBT is turned on. (d) TEM images of DS-TFET after silicide process. (c) Dopant-segregated source tunnel junction and (e) drain underlap region can be seen. (f) **Measured** transfer curves and (inset) schematic source dopant profile of pDS-TFETs. (g) **Measured** transfer curves of pDS-TFETs w/ and w/o drain underlap region.

proposed and experimentally demonstrated monolithically integrated with standard MOSFET on 300mm Si wafer.

## A. *Standard salicide process for source tunnel junction engineering and Self-aligned drain underlap design*

The dopant segregation (DS) technology can be utilized to break the implantation limitation to form a sharp source doping profile with high concentration which has already been proved in nanowire TFETs by research institute [13]. Considering the limitation of using new materials in standard Si CMOS platform, for the first time, we utilize the standard salicide process from baseline BEOL procedure to realize the dopant-segregated tunnel junction for the TFET performance enhancement, and the implantation before silicide approach was adopted in this work. In order to obtain the optimal ion implantation condition for dopant segregation technology, the NiSi/Si junctions based on baseline BEOL process are fabricated. TEM and SIMS results of NiSi/Si junction show that the segregation dopants can be accumulated at NiSi/Si interface with large concentration and steep gradient at the tail of the dopant profile (Fig. 2).

To utilize the above dopant-segregated effect in TFET for performance enhancement, a self-aligned segregation source junction is designed on standard CMOS platform (Fig. 3). NiSi is formed at the source side to obtain pocket layer at the NiSi/Si junction along the direction of the channel, and the pocket layer needs to be placed at the gate edge to obtain a large electric field for high band-to-band tunneling (BTBT) generation rate. In this work, by utilizing standard thin offset spacer and salicide process of CMOS technology, a self-aligned dopant segregated source junction can be realized right under the gate edge (Fig. 3c). The highly doped region at the silicide junction can lead to strong band bending,

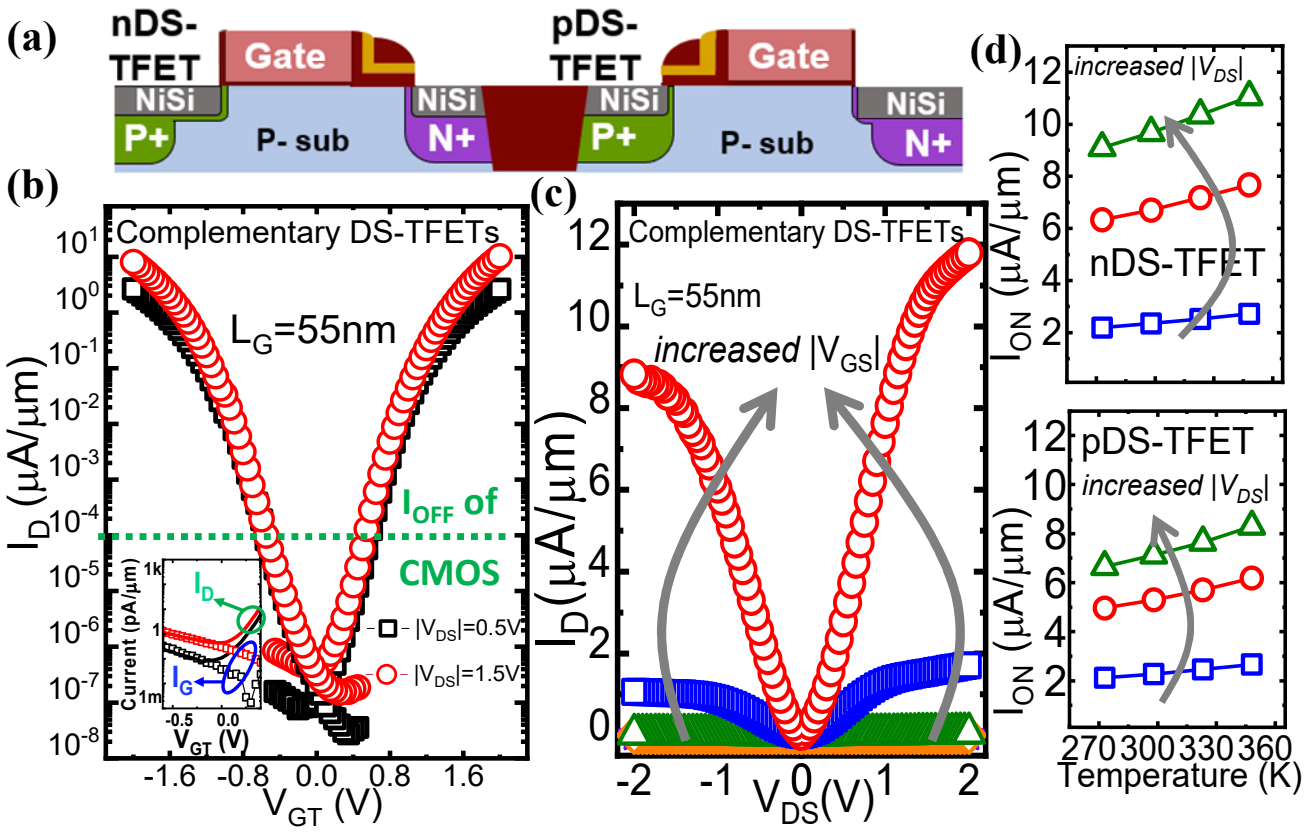

Fig.4 (a) Schematic structures of C-DS-TFETs; **Measured** (b) transfer curves and (c) output curves of fabricated C-DS-TFETs on the same wafer based on standard CMOS platform. (d) The **measured** positive $I_{ON}$ dependence on temperature confirms the band-to-band tunneling current.

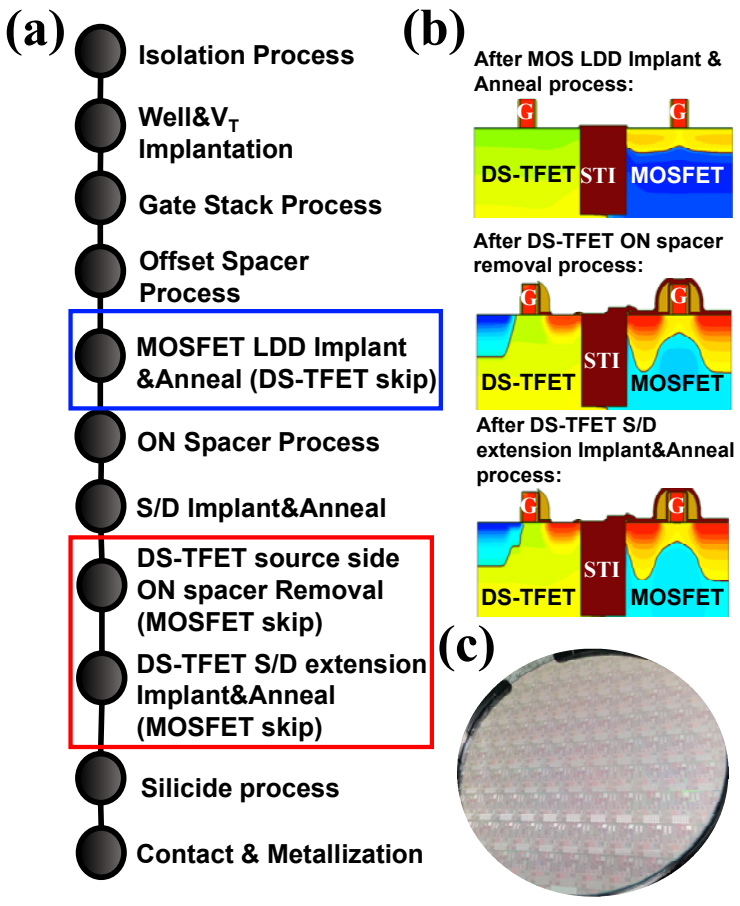

Fig.5 (a) Process flow for co-integrating C-DS-TFETs and CMOS. (b) Simulated structures by TCAD tools. (c) Photo of 300 mm wafer with C-DS-TFETs.

which will increase the BTBT generation area with high tunneling probability.

Fig. 3f shows impacts of the dopant segregation effect on TFET device performance. Compared with the conventional device with implantation junction, it can be seen that dopant segregation can significantly improve the $I_{ON}$ characteristics of TFET. Different ion implantation conditions (IMP I, II, III as shown in Fig. 3f) for dopant segregation technology are also performed and compared. The larger implantation depth results in the worse dopant segregation effects, which will degrade to the conventional implantation junction situation. With the well-designed process conditions, the device $I_{ON}$ can be significantly enhanced by 3 decades of current.

To suppress the ambipolar current of TFET, the standard thick ON spacer of CMOS technology is remained as a hard mask to form a self-aligned drain underlap region for both silicide and ion implantation (Fig. 3e). The experimental results show that the ambipolar current can be successfully suppressed (Fig. 3g).

## B. *N-type and P-type Tunneling Devices*

Currently, most of reported high-performance TFETs by industry-manufacturers are n-type devices based on III-V materials [14-19]. For III-V TFETs, there exists an inherent optimization conflict of the source/channel heterojunction between n-type and p-type TFETs [7]. For Si TFETs, the

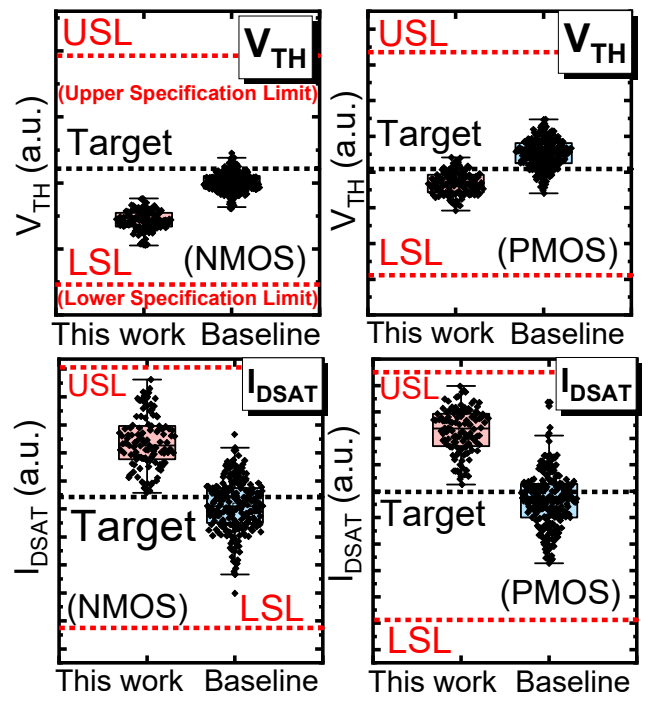

Fig.6 Comparison of performances between baseline CMOS (Baseline) and CMOS integrated with DS-TFETs (This work) based on standard CMOS platform.

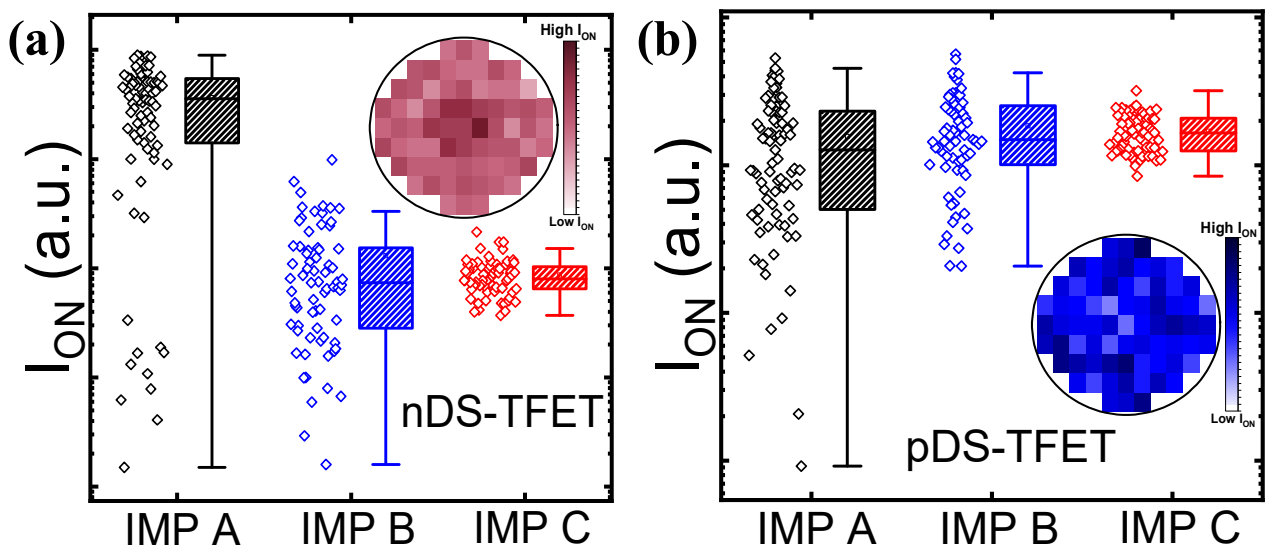

Fig.7 **Measured** $I_{ON}$ distributions in (a) nDS-TFETs and (b) pDS-TFETs. Insets are 2D mapping of measured $I_{ON}$ in optimized DS-TFETs with proposed implantation (IMP C).

different diffusion capability of opposite types of dopants also bring difficulties in fabricating complementary TFETs as shown in our previous work [8]

In this work of proposed salicide-enhanced Si TFET technology (Fig. 4a), n- and p-type DS-TFETs show the similar drive current capability with both large $I_{ON}$ of almost 10μA/μm, which is even larger than recently reported III-V TFETs [6] (Fig. 4b). Meanwhile, the $I_{OFF}$ is limited at gate leakage current level and can be 2 decades lower than the standard CMOS with the same gate length. The output curves with weak super-linear onset behavior (Fig. 4c) and the positive $I_{ON}$ dependence on temperature (Fig. 4d) further confirm the tunneling current rather than Schottky current.

## III. Monolithically Integration with Standard CMOS

To monolithically integrate DS-TFET of asymmetric sidewall structure with standard CMOS, the source ON spacer of DS-TFET needs to be removed after CMOS FEOL process and before BEOL salicide process, followed with the well-designed source/drain extension implantation and annealing process for DS-TFET (Fig. 5). Compared with CMOS, only a few masks and processes are added or modified for complementary DS-TFETs without the need of new materials, allowing simple integration with ultra-low cost. It is shown that there is no performance penalty of CMOS by utilizing the proposed fabrication process (Fig. 6).

According to our previous work [8], in standard TFETs, the source doping gradient is the dominant variation source due to its direct effect on BTBT junction area and may induce a trade-off between device performance and variability. For DS-TFET, the introduced salicide process for source tunnel junction becomes the new dominant variation source. Therefore, steep source gradient will result in large $I_{ON}$ distribution on 300mm-wafer. As shown in Fig.7, by further reasonable implantation optimization, the

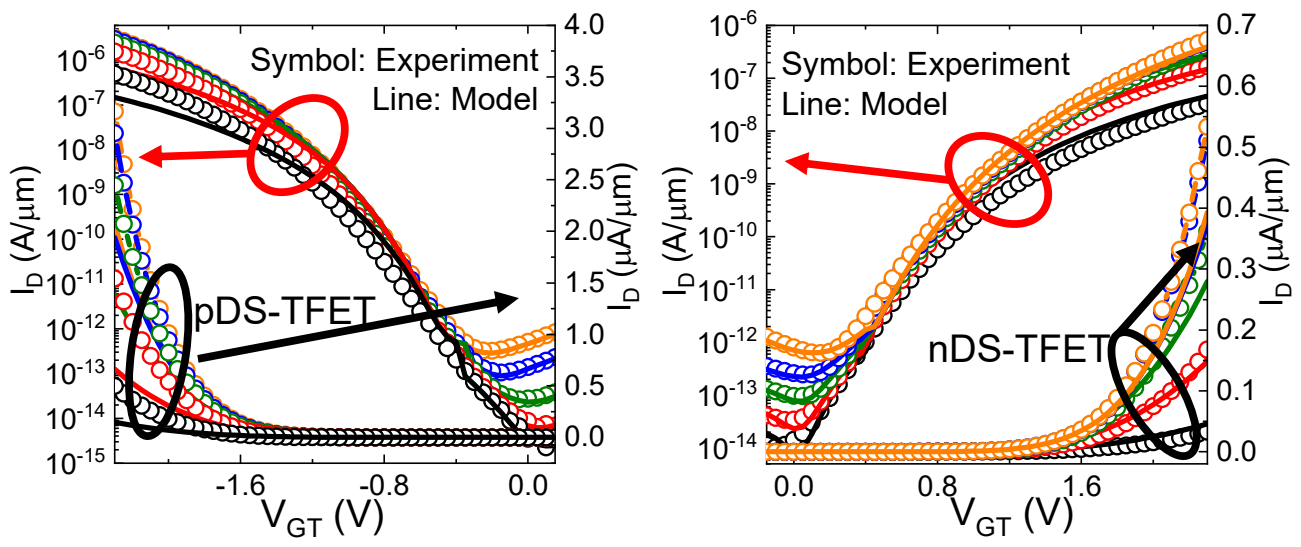

Fig.8 Good agreement between experiment (Symbol) and simulation (Line) with different $V_{DS}$.

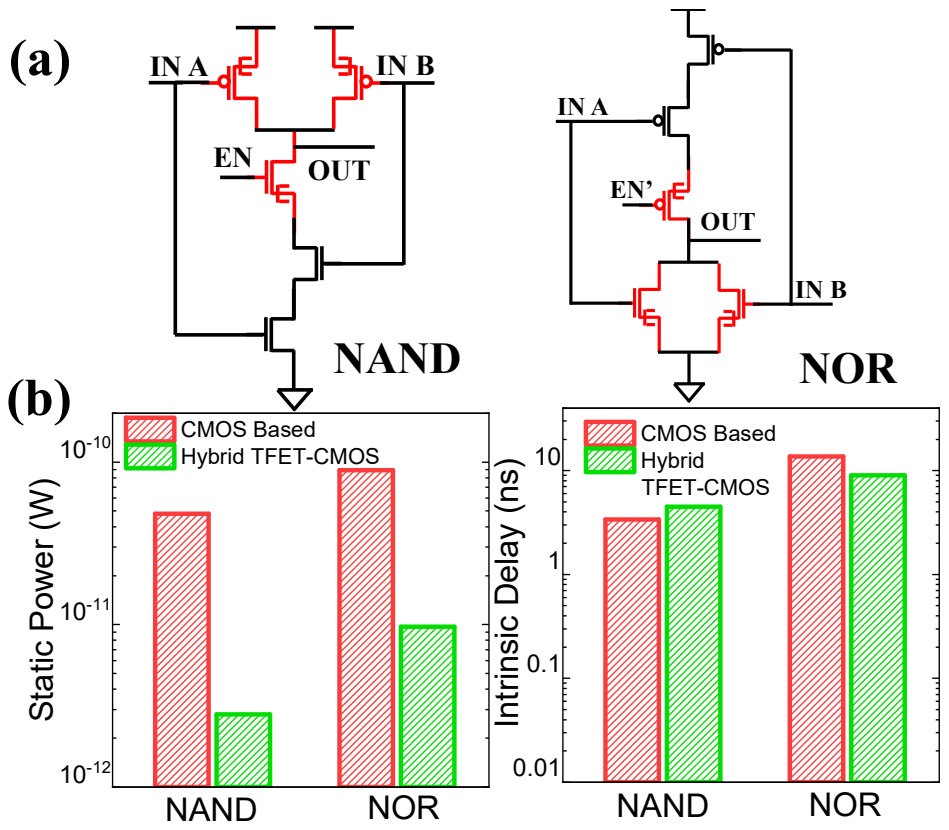

Fig.9 (a) Schematic of DS-TFET-based logic cells. (b) Static power and Intrinsic delay comparison among CMOS-based logic cells, and Hybrid TFET-CMOS topologies.

device variation can be significantly reduced while still maintaining the $I_{ON}$ superiority from the dopant segregation effect compared with conventional TFETs.

## IV. Circuit-Level Evaluation for AIoT Application

Based on the above monolithically TFET-CMOS hybrid integration platform, we further evaluate the TFET circuit implementation for AIoT. A compact model is developed [20] and calibrated with experimental results in this work (Fig. 8).

In this work, the monolithically integrated TFET and CMOS on the same water makes it possible to take advantage of low power TFET and high performance CMOS simultaneously, which is beneficial for low power and real time requirements of AIoT applications. The static power and intrinsic delay of standard logic cells are demonstrated based on our previous proposed hybrid TFET-MOSFET topologies [12] which can solve the severe current degradation issue induced by TFET stacking in circuits. Fig.9 shows that compared with CMOS logic cells, the hybrid TFET-MOSFET topologies have the much lower static power consumption for almost 1 decades with comparable intrinsic delay.

SRAM, with fast access speed, can be utilized as the weight memory for AIoT applications, while suffering from large static power consumption. Although the low $I_{OFF}$ characteristic makes TFET a promising candidate to implement low-power SRAM, its nonideal characteristics, such as unidirectional conductivity, positive P-I-N current and superlinear onset prevent it from being used in traditional 6T cell topology. Our previous work [10] has proposed 10T TFET SRAM topology, which utilizes the read-write separation and the combinational access topology to solve the problem brought by unidirectional conductivity and positive P-I-N current. However, since it introduces

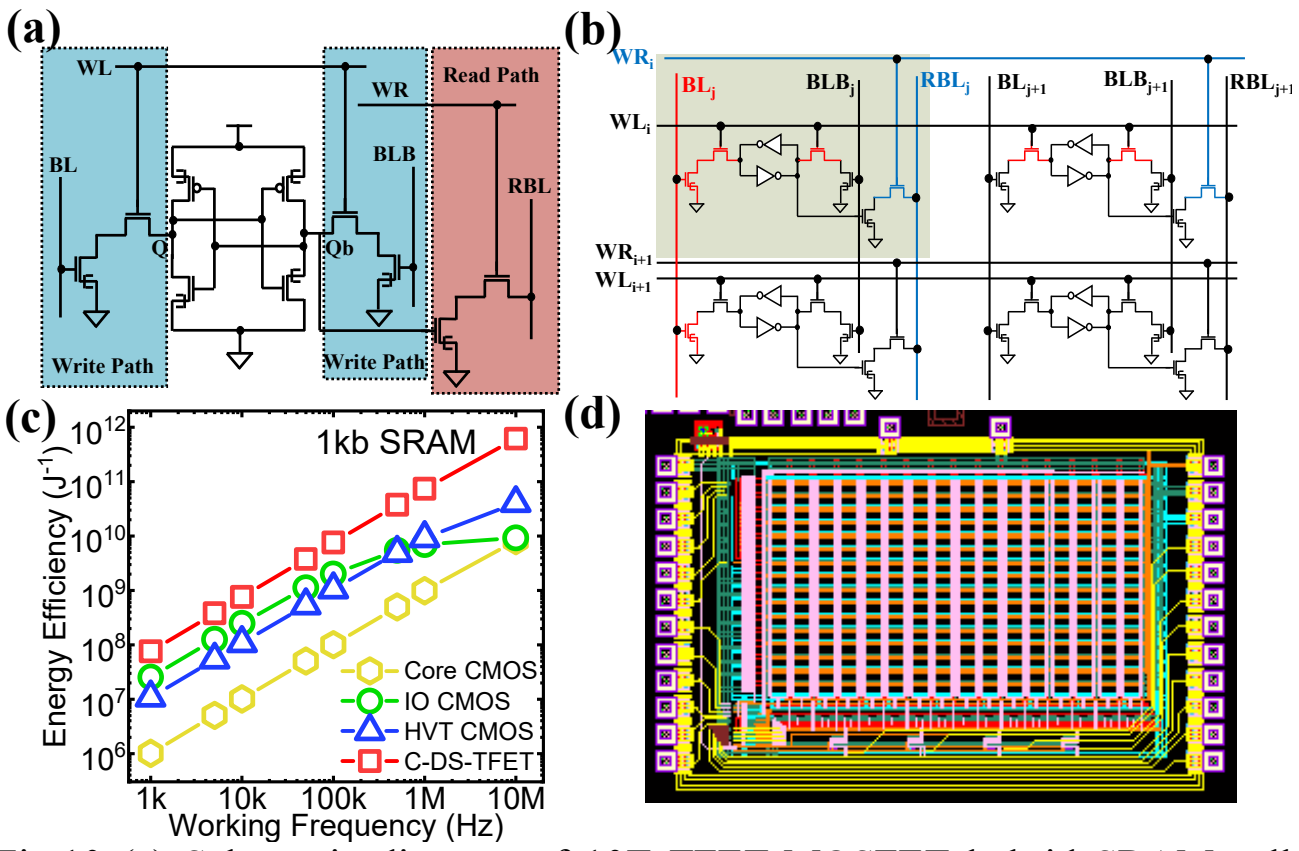

Fig.10 (a) Schematic diagram of 10T TFET-MOSFET hybrid SRAM cell topology; (b) Schematic diagram of 2x2 array of 10T TFET-MOSFET hybrid SRAM; (c) Energy efficiency comparison among DS-TFETs, core CMOS, IO CMOS and HVT CMOS; (d) Layout image of 10T TFET-MOSFET hybrid SRAM array.

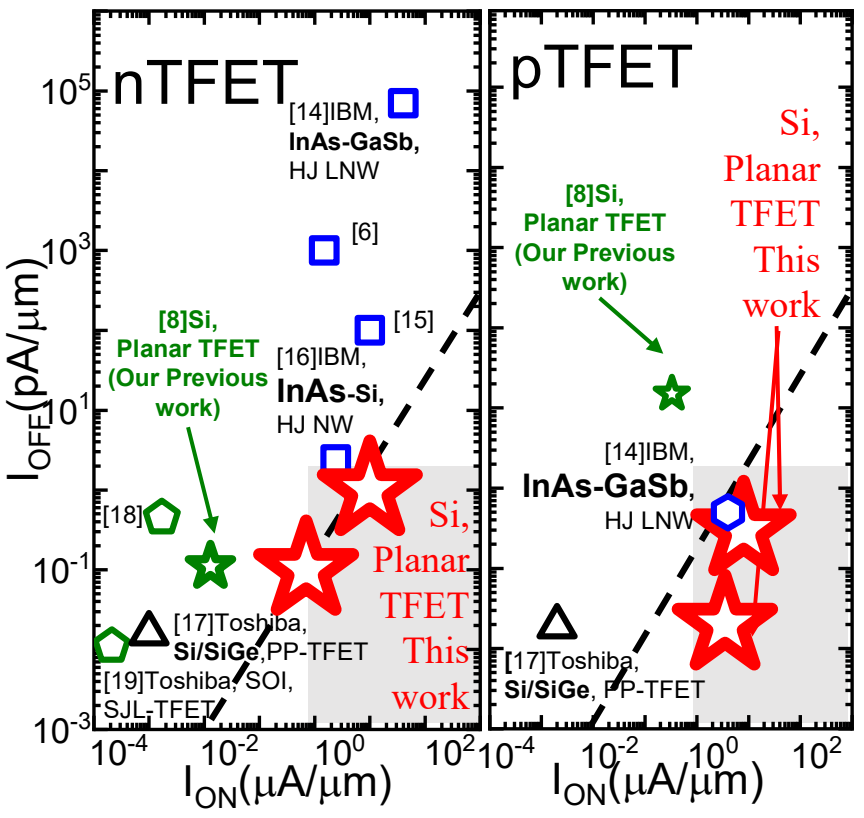

Fig.11 Performance comparison of fabricated TFETs by industry-manufacturers. The dashed line indicates $I_{ON}/I_{OFF}$ equals to $10^7$.

TFET with cascade connection in the read and write path, the series current attenuation will bring about the problem of area increase and access time degradation.

In this work, we further propose a TFET-MOSFET hybrid SRAM cell topology (Fig. 10a). It changes the series TFET structure in the access path to a TFET-MOSFET hybrid series structure and uses the same access method as 10T TFET SRAM topology [10]. Therefore, the hybrid TFET-MOSFET SRAM topology shows the faster access time. Since the standby SNM is determined by the cross-coupled inverters and the standby power is determined by the leakage path, the hybrid structure has the same noise margin and standby power as the 10T TFET-SRAM topology.

Moreover, the proposed hybrid TFET-MOSFET SRAM can solve the crosstalk issue of TFET SRAM array (Fig. 10b). Taking a 2x2 array as an example, for the write operation, there exists a device under the off-state in the TFET-MOSFET series write path of the half-selected cell. Therefore, the storage state of the half-selected cell will be not affected. For the read process, the column half-selected cell RBLj+1 is not pre-charged. Therefore, there is no current flowing through the read path. As for the row half-selected cell, the read access gate is closed by WRi+1. Therefore, the stored information is not affected as well.

The energy efficiency of hybrid TFET-MOSFET SRAM is further evaluated based on a 1kb array. It is shown that the proposed hybrid TFET-MOSFET SRAM based on DS-TFET can achieve the higher energy efficiency even than High-$V_{TH}$ (HVT) and IO CMOS based SRAM of the same technology node (Fig. 10c), which is very promising for ultralow-power AIoT applications.

## V. Conclusion

For the first time, the foundry technology of integrating the proposed complementary DS-TFET on CMOS baseline platform are presented, as well as the novel TFET-CMOS hybrid logic cell and SRAM topologies. Without CMOS performance penalty, the n- and p-type Si DS-TFETs are manufactured together with standard MOSFETs, showing the significant performance enhancement with high $I_{ON}$ and record $I_{ON}/I_{OFF}$ ratio among TFETs by industry-manufacturers (Fig. 11). Circuit-level energy saving further indicate the great potential of this novel TFET-CMOS hybrid platform for cutting-edge power-dieting AIoT.

## Acknowledgment

The authors would gratefully acknowledge SMNC and SMIC for the assistance in the device fabrication. This work was supported by National Key R&D Program of China (2018YFB2202800), NSFC (61851401, 61822401, 61927901), Beijing Nova Program of Science and Technology (Z191100001119101), BJSAMT Project (SAMT-BD-KT-22030101) and 111 Project (B18001).